# DISTRIBUTION OF MAXIMAL CLIQUE SIZE OF THE VERTICES FOR THEORETICAL SMALL-WORLD NETWORKS AND REAL-WORLD NETWORKS

Natarajan Meghanathan

Jackson State University, 1400 Lynch St, Jackson, MS, USA
natarajan.meghanathan@jsums.edu

## ABSTRACT

*Our primary objective in this paper is to study the distribution of the maximal clique size of the vertices in complex networks. We define the maximal clique size for a vertex as the maximum size of the clique that the vertex is part of and such a clique need not be the maximum size clique for the entire network. We determine the maximal clique size of the vertices using a modified version of a branch-and-bound based exact algorithm that has been originally proposed to determine the maximum size clique for an entire network graph. We then run this algorithm on two categories of complex networks: One category of networks capture the evolution of small-world networks from regular network (according to the well-known Watts-Strogatz model) and their subsequent evolution to random networks; we show that the distribution of the maximal clique size of the vertices follows a Poisson-style distribution at different stages of the evolution of the small-world network to a random network; on the other hand, the maximal clique size of the vertices is observed to be in-variant and to be very close to that of the maximum clique size for the entire network graph as the regular network is transformed to a small-world network. The second category of complex networks studied are real-world networks (ranging from random networks to scale-free networks) and we observe the maximal clique size of the vertices in five of the six real-world networks to follow a Poisson-style distribution. In addition to the above case studies, we also analyze the correlation between the maximal clique size and clustering coefficient as well as analyze the assortativity index of the vertices with respect to maximal clique size and node degree.*

## 1. INTRODUCTION

Network Science is an emerging area of research interest to study complex real-world networks from a graph theoretic point of view. We abstract the complex network as a graph with the nodes representing the vertices and the connections between any two nodes in the network modeled as edges in the graph. It is imperative that the algorithms run on these large scale graphs be as efficient as possible and do not take significant time to determine the metrics of interest. Though there exists efficient polynomial-time algorithms to determine widely studied metrics [1] like centrality, diameter, clustering coefficient, etc on these graphs, there still exists certain metrics like clique such that the problem of determining a maximum size clique is NP-hard [2]. A clique on a graph is a subset of the vertices such that there exists an edge between any two vertices in this subset; an algorithm to find cliques of various sizes (constituent nodes) could be used to identify closely-knit communities [3-5] of various sizes in complex network graphs, including both real-world networks as well as networks that evolve from theoretical models.

The "maximum size clique" for a graph of $n$ vertices is a clique of the largest size $k$ ($k \leq n$) such that there does not exist a clique of size $k + 1$ in the graph. A "maximal size clique for a vertex $i$" in a graph is the clique of the largest size that involves vertex $i$ as one of the constituent vertices. While the maximum size clique for a graph is the maximal size clique for its constituent vertices, there could exist several other vertices in the graph for which the maximal size clique is smaller than the maximum size clique. Most of the research focus in the literature is to develop exact algorithms that could determine the maximum size clique for the entire graph as efficiently as possible with respect to both time and space complexity. Very little attention has been given to determine the maximal size cliques for the individual vertices in the graph. Specifically, to the best of our knowledge, no attempt has been made to analyze the distribution of the maximal clique sizes of the individual vertices in complex network graphs. In this paper, we choose a recently proposed exact algorithm [6] to determine the size of the maximum clique for large-scale complex network graphs and extend it to determine the size of the maximal clique that a particular node is part of. Using the exact algorithm to determine maximal clique size for the individual vertices of the graphs, we determine the distribution of the maximal clique size for two categories of complex networks: The first category of complex networks correspond to networks that evolve during the transformation of a regular network to a small-world network and further to a random network; we use the well-known Watts-Strogatz model [18] to simulate the evolution of the small-world networks and random network from a regular network. The second category of complex networks are six real-world network graphs (ranging from random networks to scale-free networks). As the networks evolve from a regular network to a small-world network, we observe the maximal clique size of the vertices to be almost identical to each other and as well correspond to the maximum clique size for the entire network graph; on the other hand, as a small-world network evolves to a random network, we observe the maximal clique size of the vertices to exhibit a Poisson-style distribution. Likewise, we observe five of the six real-world network graphs (irrespective of their number of nodes and degree distribution) to exhibit a Poisson-style distribution for the maximal clique size. The above observations are significant to the study of cliques and their associated phenomenon (community detection, homophily, etc) in complex networks and such results have not been hitherto reported in the literature.

The second half of our paper focuses on identifying a computationally-light metric for the individual nodes of a graph that correlates well (either positively or negatively) to that of the maximal clique size (which we categorize as a computationally-hard metric, owing to the NP-hard nature of the problem to determine this metric and the significant time complexity involved in the exact algorithms for this metric). Once we identify such a computationally-light metric that correlates well with the maximal clique size of the vertices in complex network graphs, we could infer a ranking of the vertices based on this computationally-light metric as a ranking of the vertices based on the maximal clique size. To the best of our knowledge, we have not come across any such study to identify a computationally-light metric that correlates well with the maximal clique size for real-world network graphs. Ours is the first attempt in this direction. The two candidate computationally-light metrics that we consider are the clustering coefficient and the node degree. The clustering coefficient of a vertex is the ratio of the number of edges between the neighbors of the vertex to that of the maximum number of edges possible between the neighbors of the vertex. Our conjecture is that nodes that are part of a larger clique are more likely to have a larger clustering coefficient and vice-versa. Similarly, we conjecture that nodes that have a larger degree (number of neighbors) are likely to be part of cliques of larger size and vice-versa. Results of our correlation studies on real-world network graphs reveal that the maximal clique size has good correlation with node degree (especially as the variation in the node degree increases), whereas the maximal clique size correlates poorly with the clustering coefficient. We further confirm the positive correlation between the maximal clique size and node degree through an analysis of the Assortativity index of the vertices [1] in the real-world network graphs with respect to these two metrics. We observe the real-world network graphs could be ranked in a similar order in the decreasing order of the Assortativity index of the vertices with respect to both the maximal clique size and the node degree.

The rest of the paper is organized as follows: Section 2 describes related work on analysis of complex network graphs using cliques. Section 3 describes an efficient exact algorithm to determine the maximum clique size for an entire graph and our extension to determine the maximal clique size for the individual vertices of the graph. Section 4 presents the evolution of small-world networks and their transformation to a random network under the Watts-Strogatz model and describes the results of the maximal clique size of the networks that evolve during this transformation. Section 5 presents the real-world network graphs studied in this paper and an analysis of their degree distribution and distribution of the maximal clique size of the vertices. Section 6 presents the results of the correlation studies between the maximal clique size and clustering coefficient. Section 7 presents the results of the correlation studies between the maximal clique size and the node degree. Section 8 presents the results of Assortativity index-based analysis of the real-world network graphs with respect to maximal clique size and node degree. Section 9 concludes the paper. Throughout the paper, we use the terms 'node' and 'vertex', 'link' and 'edge' interchangeably. They mean the same.

## 2. RELATED WORK

The research focus with regards to cliques in the context of complex networks is to come up with efficient heuristics to reduce the run-time complexity in determining the maximum size clique for the entire network graph. Though branch-and-bound has been the common theme among these works, the variation is in the approach used to arrive at the bounds and enforce them in the search space. Strategies used for pruning the search space are typically based on node degree (e.g., [6]), vertex ordering (e.g., [7]) and vertex coloring (e.g., [8]). Recently, a parallelized approach [9] for branch and bound has also been proposed for determining cliques in real-world networks ranging from 1000 to 100 million nodes. Nevertheless, none of the research so far has focused on identifying correlation between the maximal clique size for an individual vertex (the size of the largest clique that a particular vertex is part of) with any of the commonly studied metrics (like node degree, clustering coefficient) for network analysis. Ours is the first step in this direction. With the problem of determining maximum size clique for the entire network graph and maximal size cliques for the individual vertices being NP-hard and computationally time-consuming for complex real-world networks of larger size, it becomes imperative to analyze the correlations of the maximal clique size values of the individual vertices with that of the network metrics that can be easily computed so that meaningful inferences about maximal clique size values can be made.

## 3. CLIQUE

A clique is a sub graph of a graph in which all the vertices are adjacent to each other. The problems of finding maximum size clique for the entire graph as well as the maximal size cliques for the individual nodes are NP-hard problems [2]. Several exact algorithms (that at the worst case incur exponential time for a NP-hard problem) have been proposed to determine maximum size cliques for sparse graphs. Recently, with the surge in interest to analyze large real-world networks from a graph theoretic point of view, researchers have proposed efficient exact algorithms (e.g., [6-9]) to determine maximum size cliques for large/dense graphs. The common theme [10] behind these algorithms is a branch and bound approach of searching through all possible candidate cliques and limiting the search to only viable candidate sets of vertices whose agglomeration has scope of being a clique of size larger than the currently known clique found as part of the search; the variation among these exact algorithms is the pruning strategy (the approach taken to compute the bounds and use them) to limit the search. In this section, we will describe one such branch and bound-technique based exact algorithm that has been recently proposed in the literature [6] to determine maximum size clique in large network graphs and explain our modification to the algorithm so that it can be used to determine the maximal cliques that each vertex in the graph is part of; the largest among these cliques is the maximum size clique for the entire graph.

Figure 1 outlines the pseudo code of the algorithm (proposed originally in [6]) to determine the maximum size clique for an entire graph. The algorithm starts with an estimate of 0 for the maximum size clique (variable *max*) in the entire graph; the value for *max* is updated as and when a clique of size larger than the latest value of *max* is found. The procedure MAXCLIQUE proceeds in iterations, with each iteration designed to determine the maximum size clique for the entire graph that could also include vertex $v_i$ (considered in the increasing order of the IDs). In a particular iteration, vertex $v_i$ is considered worthy of exploration for presence in a maximum size clique only if its degree is at least the value of *max* at that time (i.e., only vertices that could be part of a clique of size larger than the currently known maximum size clique are considered - a pruning strategy). For each such vertex $v_i$, a candidate set $U$ of neighbor vertices $v_j$ (whose degree is at least the latest value for *max*) is constructed and passed to the sub routine CLIQUE to find a clique among these vertices; the initial size of the clique is 1 - accounting for $v_i$.

---

**Procedure** MAXCLIQUE ($G = (V, E)$)
   $max \leftarrow 0$
   **for** $i$ : 1 to |V| **do**
     **if** degree($v_i$) $\geq$ *max* **then**
       $U \leftarrow \phi$
       **for** each $v_j \in$ Neighbor($v_i$) **do**
         **if** degree($v_j$) $\geq$ *max* **then**
           $U \leftarrow U \cup \{v_j\}$
       CLIQUE($G, U, 1$)

**Subroutine** CLIQUE($G = (V, E)$, $U$, *size*)
   // *size* is the size of clique found so far
   **if** $U = \phi$ **then**
     **if** *size* > *max* **then**
       *max* $\leftarrow$ *size*
     **return**
   **while** |U| > 0 **do**
     **if** *size* + |U| $\leq$ *max* **then**
       **return**
     select any vertex $u$ from $U$
     $U \leftarrow U \setminus \{u\}$
     $N'(u) := \{w \mid w \in$ Neighbor($u$) $\wedge$
                   degree($u$) $\geq$ *max*$\}$
     Clique($G, U \cap N'(u),$ *size* + 1)

---

Figure 1. Exact Algorithm to Determine Maximum Size Clique for a Graph (adapted from [6])

The sub routine CLIQUE called with vertex $v_i$ as the first constituent vertex of the largest possible clique involving $v_i$, expands with one vertex at a time through a combination of iterations and recursions; the sub routine runs as long as the size of the set $U$ is greater than zero or if the current value of *max* is less than the sum of the sizes of the set $U$ and the current clique found so far (a pruning strategy). In each such iteration, a vertex $u$ (that is also a neighbor of the starting vertex $v_i$ and the other vertices in the clique determined so far) is randomly removed from the set $U$ and the neighbors of $u$ that are also present in $U$ (and hence are neighbors of the starting vertex $v_i$ and the other vertices that are part of the clique found so far) are only further considered to be candidates that could be part of the clique, and a recursive call to the CLIQUE sub routine is made with the value of variable *size* (the size of the largest clique found so far involving vertex $v_i$) incremented by 1 - accounting for $u$ as the latest entrant in the clique determined so far. Each recursive call to CLIQUE is accompanied by an iteration where a vertex $u$ (that is also a neighbor of the vertices already part of the clique) is removed from the set $U$ passed to the sub routine and only the neighbors of $u$ that are also neighbors of the vertices already in the clique are considered. During any such recursive call, if the size of the set $U$ passed to the sub routine CLIQUE reaches zero, the algorithm terminates the sequence of recursions and updates the value of *max* if the size of the clique determined so far involving vertex $v_i$ is larger than the current value of *max*. During the sequence of returns from the recursive calls, it is possible that a new sequence of recursions and iterations is triggered due to the presence of a neighbor $u$ of $v_i$ that has scope for being in a clique (involving $v_i$) of size larger than the clique found so far for the entire graph. The algorithm explores all such possible cliques involving vertex $v_i$ that have scope for exceeding the currently known maximum size clique for the entire graph.

At the end, the algorithm returns the maximum size clique for the entire graph that also happens to be the maximal size clique involving some vertex $v_i$ such that there is no other vertex $v_j$ ($i > j$) that is also part of the clique. Since the algorithm proceeds with vertices in the increasing order of their IDs, if the maximum size clique for the entire graph involves at least one vertex $v_i$ with a smaller ID, the presence of the maximum size clique is detected much earlier and the subsequent iterations (with vertices whose IDs are greater than $v_i$, but could be part of only cliques of size smaller or equal to the maximum size clique of the entire graph involving $v_i$) are merely pruned, contributing to the time-efficiency of the algorithm. Hence, the labeling of the vertices with their IDs plays a significant role in the run-time complexity of the algorithm; the algorithm is capable of quickly determining the maximum size clique if the latter comprises of at least one vertex with a smaller ID.

---

**Procedure** MAXIMALCLIQUE ($G = (V, E)$)
    **for** $i$ : 1 to $|V|$ **do**
        *maximalCliqueSize*[$v_i$] ← 0
        $U ← \phi$
        **for** each $v_j \in$ Neighbor($v_i$) **do**
            $U ← U \cup \{v_j\}$
        CLIQUE($G, v_i, U, 1$)

**Subroutine** CLIQUE($G = (V, E), v_i, U, size$)        // *size* is the size of clique found so far for vertex $v_i$
  **if** $U = \phi$ **then**
    **if** *size* > *maximalCliqueSize*[$v_i$] **then**
        *maximalCliqueSize*[$v_i$] ← *size*
    **return**
  **while** $|U| > 0$ **do**
    **if** *size* + $|U|$ ≤ *maximalCliqueSize*[$v_i$] **then**
        **return**
    select any vertex $u$ from $U$
    $U ← U \setminus \{u\}$
    $N'(u) := \{w \mid w \in$ Neighbor($u$) $\wedge$ degree($u$) ≥ *maximalCliqueSize*[$v_i$]$\}$
    Clique($G, v_i, U \cap N'(u), size + 1$)

---

Figure 2. Exact Algorithm to Determine the Maximal Clique Size for each Vertex in a Graph (adapted from [6])

Figure 2 illustrates our modifications (to determine the size of the maximal clique that each vertex is part of) to the pseudo code of the algorithm presented in Figure 1. The tradeoff is an increase in the run-time of the algorithm: we cannot just prune our search based on the vertex IDs; we have to explore the neighborhood of each of the vertices to determine the maximal size clique that each vertex is part of. Since to start with, the maximal size clique known for vertex $v_i$ is 0, there is no need to filter the neighbors of $v_i$ in procedure MAXIMALCLIQUE based on the degree of the neighbors; all neighbors of $v_i$ are included in the set $U$ and passed onto the sub routine CLIQUE. However, we could retain all of the pruning steps in sub routine CLIQUE (called to find the maximal size clique for each of the vertices $v_i$) and recursive calls to the same: there is no need to explore the neighbors of vertex $u$ whose degree is less than that of the currently known maximal clique size for vertex $v_i$.

## 4. SMALL-WORLD NETWORKS AND THEIR MAXIMAL CLIQUE SIZE DISTRIBUTION

Small-world networks are characteristic of having a smaller path length (number of hops) between any two vertices and at the same time maintaining a higher clustering coefficient, a measure of the probability of link between any two neighbor nodes of a node. The clustering coefficient for a node is formally defined as the ratio of the actual number of links between any two neighbors of the node to that of the maximum possible number of links between any two neighbors of the node. Though all classes of complex networks (small-world network, random network, scale-free network, etc) exhibit a relatively smaller path length between any two nodes in the network, it is only the small-world networks that maintain a smaller path length without incurring significant loss in the clustering coefficient of the nodes. In this section, we analyze the distribution of the maximal clique size of the vertices during the evolution of a small-world network from a regular network and the subsequent transformation of the small-world network to a random network, all of which are simulated according to the well-known Watts-Strogatz model [18], hereafter, referred to as the WS model.

Initially, we start with a regular network wherein the number of neighbors per node (i.e., the number of links per node, identified as $K_{regular}$) is the same (and typically, an even number of links per node) as well as there is a particular pattern in the distribution of the links in the network (typically dependent on the dimension of the network). In this paper, we restrict ourselves to a one-dimensional regular network. The regular network envisioned in this paper has a ring as the underlying topological structure. Each node is connected to at least two other nodes (i.e., to the two neighboring nodes that are each one hop away in the ring): if there are more than 2 links per node, then the node is connected to nodes in the increasing order of the hop count in the ring. In general, if the number of links per node is $K_{regular}$, then a node is connected to neighbor nodes that are 1, 2, ..., $K_{regular}/2$ hops away from the node on the ring. Figure 3 displays a 10-node regular network with four links per node (i.e., $K_{regular} = 4$) and each node is connected to nodes that are 1 and 2 hops away from it in the ring.

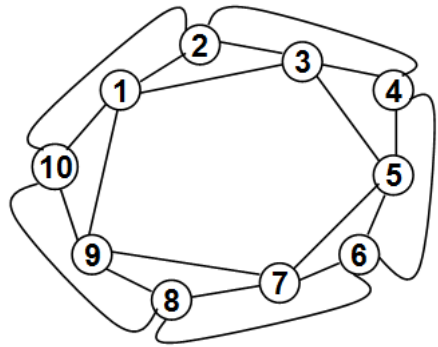

Figure 3. Example for an One-Dimensional Regular Network ($K_{regular} = 4$ Links per Node)

The WS model operates based on a tuning parameter called the probability of link rewiring ($P_{rewire}$). We rewire each link in the regular network with the probability $P_{rewire}$. As part of the procedure to decide whether or not to rewire a link *u-v*, we generate a random number (in the range 0 to 1) and if it is less than $P_{rewire}$, then we decide to rewire the link. When a link *u-v* is chosen for rewiring, we choose a target node *w* uniform-randomly among the nodes in the network (such that *w* is neither *u* nor *v*), remove the link *u-v* and connect node *u* with node *w* (i.e., add the link *u-w* to the network). We repeat the above procedure for every link in the initial regular network. Note that the newly added links are not considered for rewiring.

We conduct simulations to transform a regular network to a small-world network and subsequently to a random network according to the WS model. The simulations are conducted for networks of 100 nodes and 200 nodes; the probability of rewiring is varied from values of 0.01 to 0.1, in increments of 0.01 (referred to as small-world network zone), and from values of 0.1 to 1.0, in increments of 0.1 (referred to as random network zone). The reasoning behind the above distinction for the probability of rewiring is based on our observations from the simulation results: for $P_{rewire}$ values of 0.01 to 0.1, the average diameter of any node in the network (average of the maximum of the number of hops from a node to any other node) reduces significantly, but with only a moderate reduction in the clustering coefficient - a phenomenon characteristic of small-world networks. On the other hand, as we vary the probability of rewiring from 0.1 to 1.0, the average diameter of any node in the network reduces only marginally, whereas the clustering coefficient reduces significantly, indicating the transformation of the small-world network to a random network. We also vary the initial number of links per node ($K_{regular}$) in the regular network from 4 to 20, in increments of 2. The results presented in Figures 4-10 are the average of the results observed for 100 network graphs, simulated for each value of the number of nodes (100 and 200 nodes) and each value of the probability of rewiring as mentioned above.

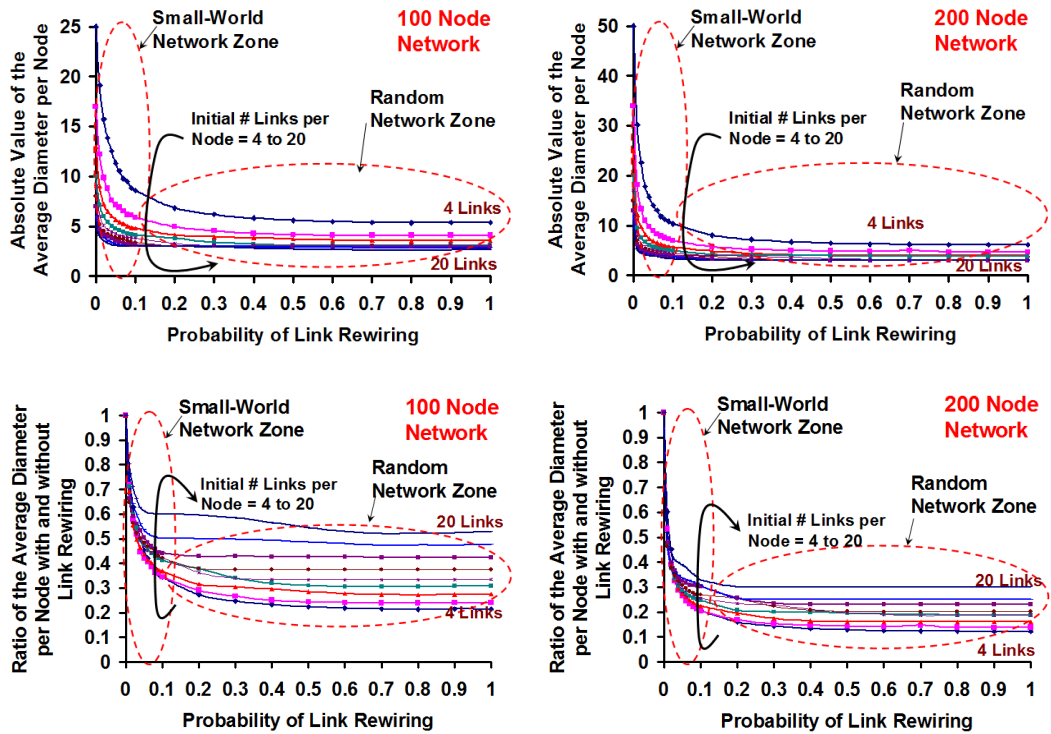

Figure 4. Impact of the Probability of Link Rewiring and the Initial Number of Links per Node on the Average Diameter per Node: Transition from Regular Network to Small-World Network and Random Network

Figure 4 captures the absolute values of the average diameter of any node in the network as well as the ratio of the average diameter with and without rewiring. For a given probability of rewiring, we observe the absolute average value for the diameter to be smaller when we start with a regular network with a larger number of links per node. As we do rewiring, within the small-world zone, we observe the differences in the average diameter per node (for different values of $K_{regular}$) to reduce significantly (in an exponential fashion); in the random network zone, the average diameter per node for different values of $K_{regular}$ does not vary appreciably. Based on the results for the ratio of the average diameter per node with and without rewiring, we observe that the percentage decrease in the average diameter per node is much higher for regular networks with fewer number of initial links, indicating the effectiveness of rewiring in reducing the path length. As we increase the number of nodes from 100 to 200, we observe the average diameter per node to further reduce (for a given probability of rewiring and initial number of links per node in the regular network), especially as we transform from a small-world network zone to a random network zone.

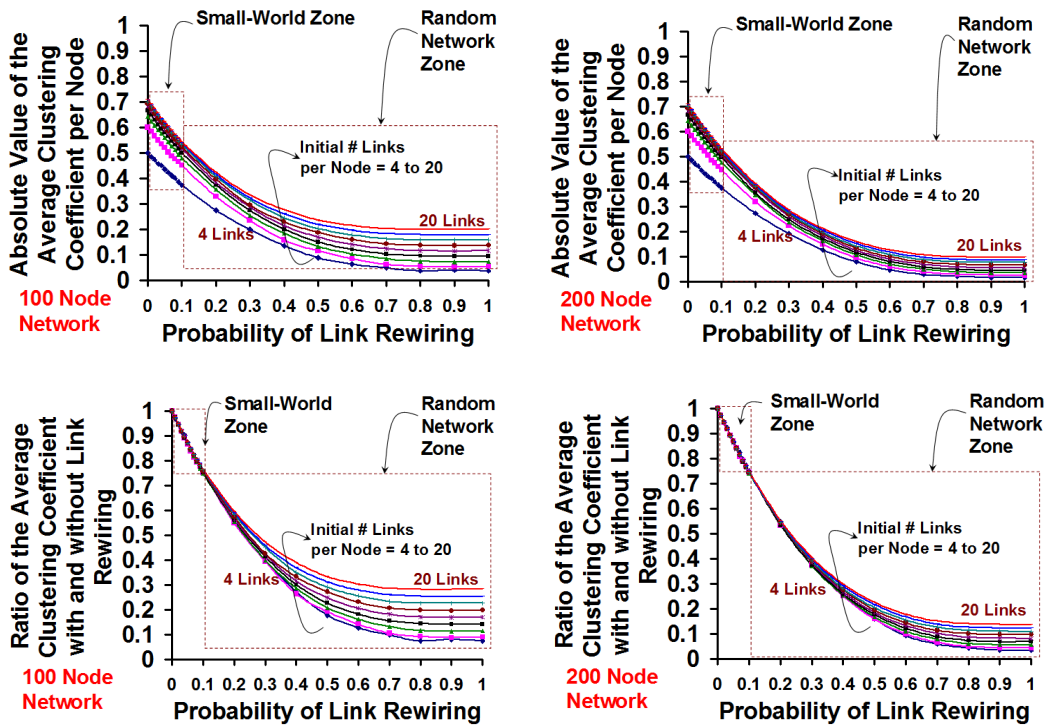

Figure 5. Impact of the Probability of Link Rewiring and the Initial Number of Links per Node on the Average Clustering Coefficient per Node: Transition from Regular Network to Small-World Network and Random Network

Figure 5 captures the reduction in the average clustering coefficient per node (average of the clustering coefficient of all the nodes in the network) due to rewiring. We observe the clustering coefficient to reduce only by about 25% in the small-world zone and the percentage reduction is the same for all values of $K_{regular}$, indicating that regular networks that start with a relatively larger value of $K_{regular}$ (expected to have a larger initial clustering coefficient, without rewiring) continue to maintain a relatively larger value (when compared to the regular networks that start with a smaller value of $K_{regular}$) for the average clustering coefficient per node with rewiring. As we enter the random network zone and with increase in the probability of rewiring, we observe the percentage decrease in the clustering coefficient to be significantly higher for random networks whose predecessor regular networks had a fewer number of initial links per node. For a given probability of rewiring and initial number of links per node in the regular network, we observe the clustering coefficient to reduce at a relatively faster rate (with respect to both the magnitude and the rate of decrease) for networks of 200 nodes compared to networks of 100 nodes. Also, in the random network zone, for a fixed probability of rewiring and number of links per node, the variation in the clustering coefficient per node for various values of $K_{regular}$ tends to reduce with increase in the total number of nodes from 200 to 100.

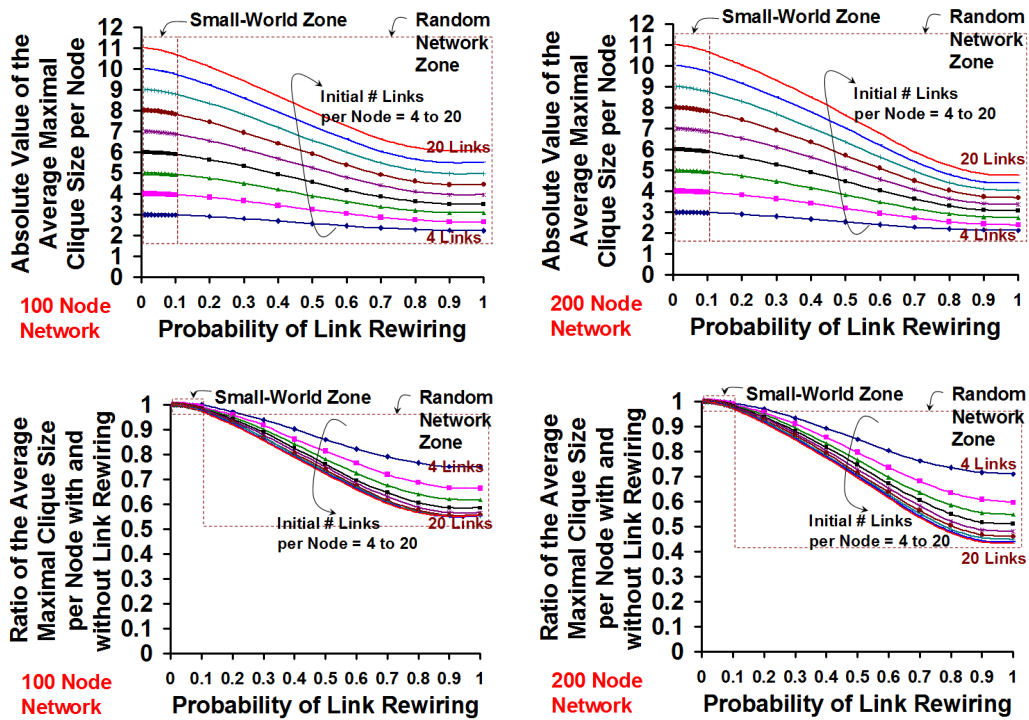

Figure 6. Impact of the Probability of Link Rewiring and the Initial Number of Links per Node on the Average Maximal Clique Size per Node: Transition from Regular Network to Small-World Network and Random Network

Figure 6 captures the variation in the average maximal clique size per node (average of the maximal clique size of all the nodes, measured at the end of rewiring) for various values of the probability of rewiring and the initial number of links per node in the originating regular network. We observe that the small-world zone does not suffer any noticeable decrease in the average maximal clique size per node and the ratio of the average maximal clique size per node with and without rewiring is close to 1. As we transition from the small-world zone to the random network zone, we observe the average maximal clique size to reduce relatively at a much faster rate, with increase in the probability of rewiring. We also notice that the rate of decrease in the average maximal clique size per node is much more steep for larger values of $K_{regular}$, indicating that cliques of larger size get dismantled due to rewiring; on the other hand, even though the absolute values for the average maximal clique size per node is much smaller for lower values of $K_{regular}$, the rate of decrease in the average maximal clique size is much flat, vindicating that there are larger cliques to start with. An interesting observation is that the average maximal clique size per nodes for different values of $K_{regular}$ tend to get closer as we increase the probability of rewiring in the random network zone, and such a converge is relatively more pronounced for networks with 200 nodes, compared to 100 nodes. Accordingly, for a given $P_{rewire}$ and $K_{regular}$, the rate of decrease in the average maximal clique size per node is much more steeper for networks with 200 nodes.

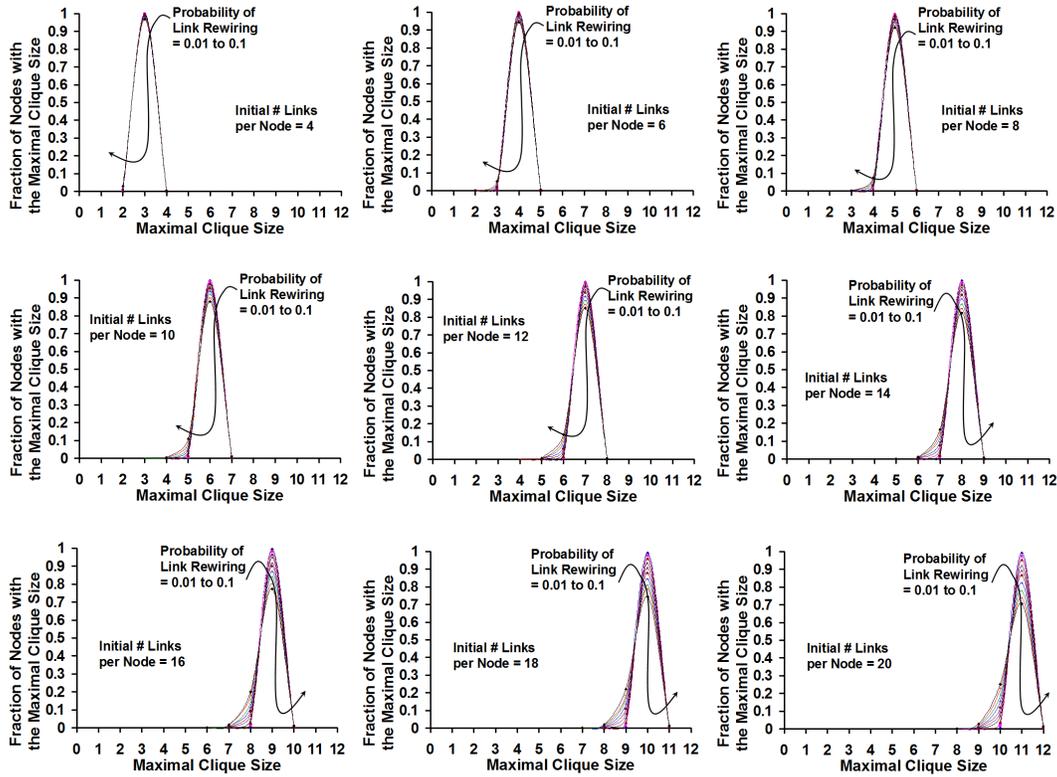

Figure 7. Distribution of the Maximal Clique Size vs. Initial Number of Links per Node: Transition from Regular Network to Small World Network [100 Node Network]

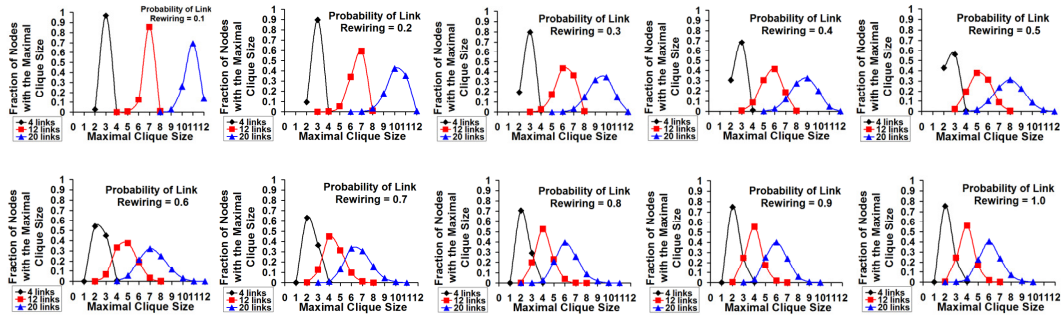

Figure 8. Distribution of the Maximal Clique Size vs. Probability of Link Rewiring: Transition from Small World Network to Random Network [100 Node Network]

Figures 7-10 capture the variation in the maximal clique size for the nodes in the small-world zones and random network zones. For a given value of $P_{rewire}$ and $K_{regular}$, we observe the distribution of the maximal clique size is Poisson for both the zones. Figures 7 and 9 capture the distribution of the maximal clique size in the small-world zone. For smaller values of $K_{regular}$ (4 and 6 links per node), we observe the maximal clique size per node to be very close to the average value for all the nodes; as we increase the value of $K_{regular}$, we observe the maximal clique size per node to vary slightly, but not much different from the average value for the maximal clique size - coinciding with the invariant nature of the average maximal clique size per node observed in Figure 6. For a given $K_{regular}$, we observe the initial value (also the average value) for the maximal clique size per node in a regular network is $1 + K_{regular}/2$; in the small-world zone, the average value for the maximal clique size per node stays much closer to this initial value for the maximal clique size per node and the variation is much minimal. For networks with larger $K_{regular}$ values, the values for the maximal clique size per node is less than the average value by at most 2 and greater than the average value by at most 1, and as observed in Figures 7 and 9, these deviations occur with a vary small probability. The Poisson curve for the maximal clique size per node shifts to the right in such a way that the peak for the curve increases by a value of 1 as we increase the value of $K_{regular}$ by 2. For a given $K_{regular}$, the tallest peak in the distribution of the maximal clique size per node is observed for a lower probability of rewiring (0.01) and the most shallow peak with relatively larger variation is observed for high probability of rewiring (0.1). Coinciding with the observations made in Figure 6, for a given value of $K_{regular}$ and $P_{rewire}$, there is not much variation in the distribution of the maximal clique size per node for networks of 100 nodes and 200 nodes.

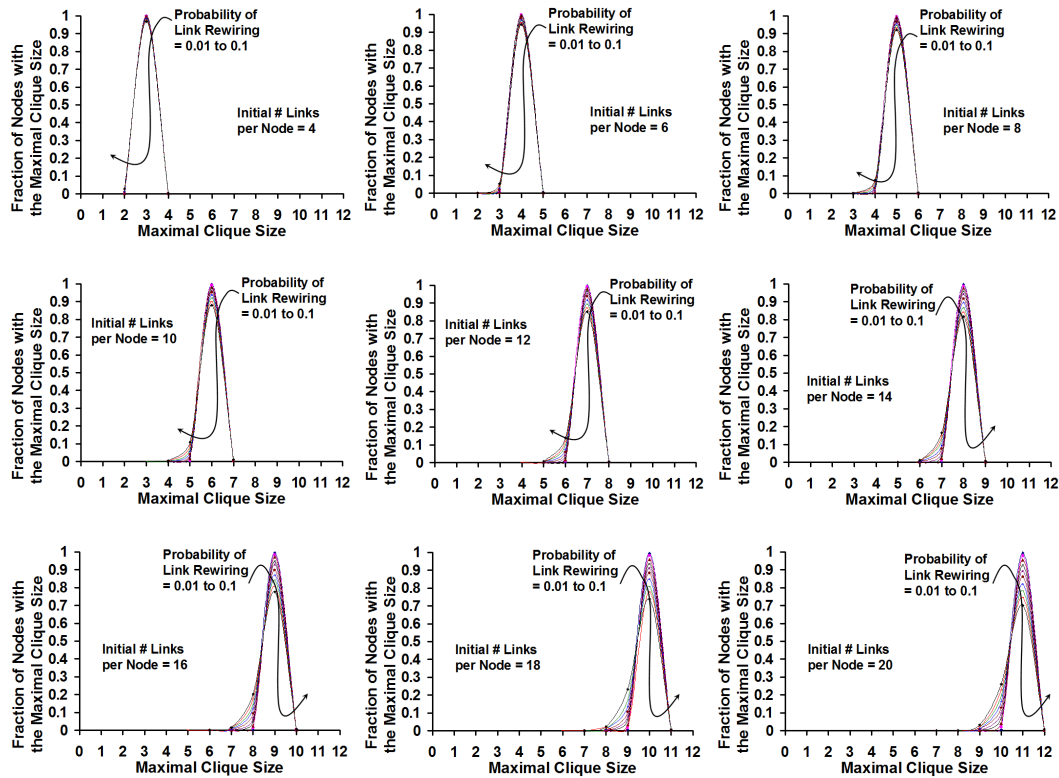

Figure 9. Distribution of the Maximal Clique Size vs. Initial Number of Links per Node: Transition from Regular Network to Small World Network [200 Node Network]

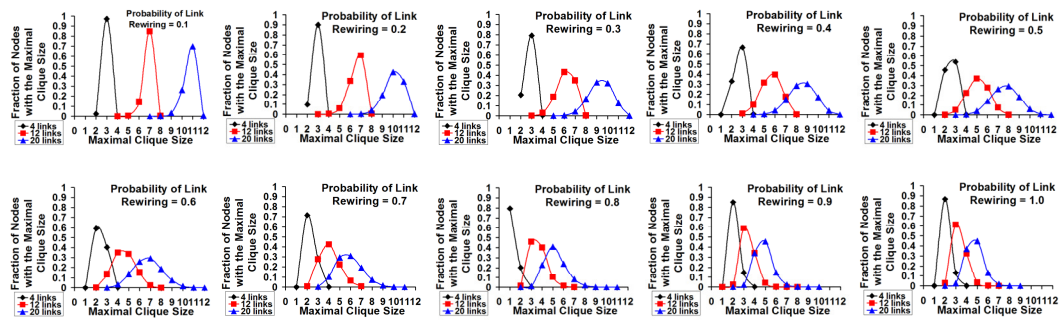

Figure 10. Distribution of the Maximal Clique Size vs. Probability of Link Rewiring: Transition from Small World Network to Random Network [200 Node Network]

Figures 8 and 10 capture the variation in the maximal clique size for the nodes in the random network zone for a given probability of rewiring and varying the initial number of links per node with values of 4, 12 and 20 links - scenarios that exhibit minimal, moderate and maximum variation in the maximal clique size per node as the probability of rewiring increases. For lower values of the probability of rewiring (0.1 and 0.2; when the network is still in the small-world zone), the distribution of the maximal clique size per node is taller for each value of $K_{regular}$ and the distributions are non overlapping (as the $K_{regular}$ values are 4, 12 and 20, the average maximal clique size is around 3, 7 and 11 - vindicating the non-overlapping nature of the peaks and the distribution of the maximal clique size for lower values of $P_{rewire}$). With increase in $P_{rewire}$, we start observing the distributions of the maximal clique size for the three fairly different values of $K_{regular}$ to start overlapping; the distributions tend to shift to the left - coinciding with a decrease in the average maximal clique value. With increase in $P_{rewire}$, the shift towards lower values of the maximal clique is more pronounced for networks with a larger $K_{regular}$ value, vindicating the rapid fall in the average maximal clique size; also for larger values of $K_{regular}$, the distributions for the maximal clique become more spread out with increase in $P_{rewire}$ - lowering the probability of finding the maximal clique size per node to be close to the average value. On the other hand, for networks with lower values of $K_{regular}$, the distribution for the maximal clique size remains fairly narrow (even with increase in the $P_{rewire}$ values), indicating that it is still possible to observe the maximal clique size for any node to be close to the average value.

## 5. REAL-WORLD NETWORK GRAPHS AND THEIR ANALYSIS

In this section, we describe the network graphs analyzed and illustrate the degree distribution and the distribution of the maximal clique size of the vertices in the network graphs. We do so to understand the topological structure of the real-world network graphs as well as to elucidate the impact of the degree and maximal clique size distribution of the vertices on the correlation between the centrality values and the maximal clique size observed for the vertices. The network graphs analyzed are briefly described as follows: (i) *Zachary's Karate Club* [11]: Social network of friendships (78 edges) between 34 members of a karate club at a US university in the 1970s; (ii) *Dolphins' Social Network* [12]: An undirected social network of frequent associations (159 edges) between 62 dolphins in a community living off Doubtful Sound, New Zealand; (iii) *US Politics Books Network* [13]: Nodes represent a total of 105 books about US politics sold by the online bookseller Amazon.com. A total of 441 edges represent frequent co-purchasing of books by the same buyers, as indicated by the "customers who bought this book also bought these other books" feature on Amazon; (iv) *Word Adjacencies Network* [14]: This is a word co-appearance network representing adjacencies of common adjective and noun in the novel "David Copperfield" by Charles Dickens. A total of 112 nodes represent the most commonly occurring adjectives and nouns in the book. A total of 425 edges connect any pair of words that occur in adjacent position in the text of the book; (v) *US College Football Network* [15]: Network represents the teams that played in the Fall 2000 season of the American Football games and their previous rivalry - nodes (115 nodes) are college teams and there is an edge (613 edges) between two nodes if and only if the corresponding teams have competed against each other earlier; (vi) *US Airports* 1997 *Network*: A network of 332 airports in the United States (as of year 1997) wherein the vertices are the airports and two airports are connected with an edge (a total of 2126 edges) if there is at least one direct flight between them in both the directions. Data for networks (i) through (v) can be obtained from http://www-personal.umich.edu/~mejn/netdata/. Data for network (vi) can be obtained from: http://vlado.fmf.uni-lj.si/pub/networks/pajek/data/gphs.htm.

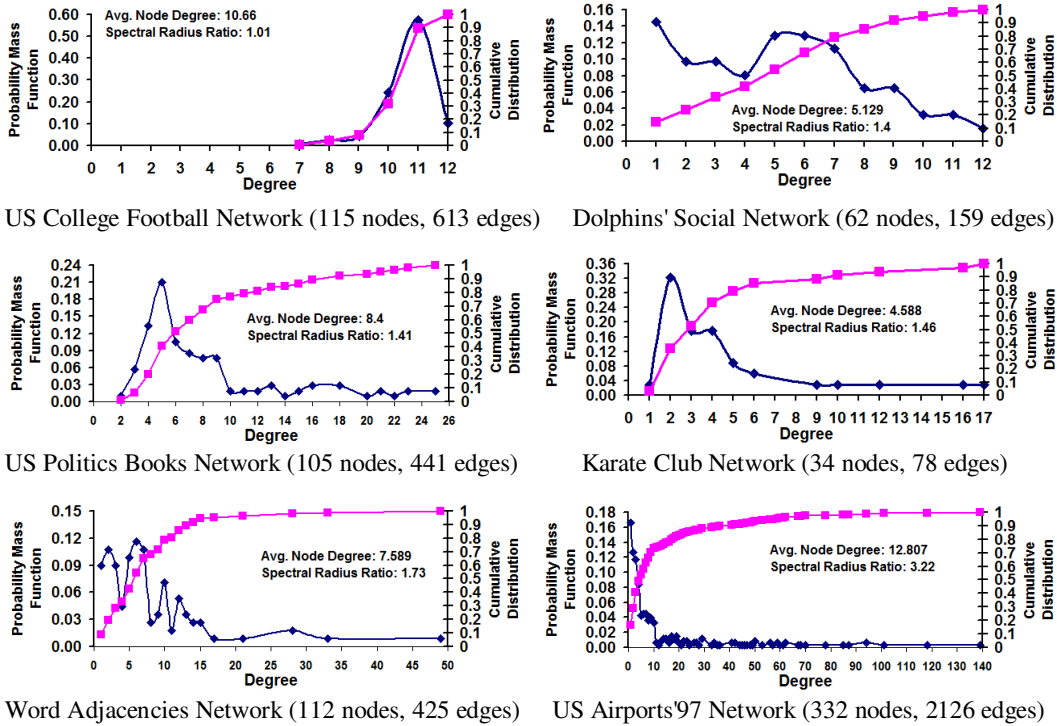

US College Football Network (115 nodes, 613 edges)   Dolphins' Social Network (62 nodes, 159 edges)

US Politics Books Network (105 nodes, 441 edges)   Karate Club Network (34 nodes, 78 edges)

Word Adjacencies Network (112 nodes, 425 edges)   US Airports'97 Network (332 nodes, 2126 edges)

Figure 11. Distribution of Node Degrees (Probability Mass Function and Cumulative Distribution)

### 5.1. Degree Distribution of the Real-World Network Graphs

Figure 11 presents the degree distribution of the vertices in the six network graphs in the form of both the Probability Mass Function (the fraction of the vertices with a particular degree) and the Cumulative Distribution Function (the sum of the fractions of the vertices with degrees less than or equal to a certain value). We also compute the average node degree and the spectral radius degree ratio (ratio of the spectral radius and the average node degree); the spectral radius (bounded below by the average node degree and bounded above by the maximum node degree) is the largest eigenvalue of the adjacency matrix of the network graph, obtained as a result of computing the eigenvector centrality of the network graphs. The spectral radius degree ratio is a measure of the variation in the node degree with respect to the average node degree; the closer the ratio is to 1, the smaller the variations in the node degree and the degrees of the vertices are closer to the average node degree (characteristic of random graph networks). The farther is the ratio from 1, the larger the variations in the node degree (characteristic of scale-free networks). Figure 11 presents the degree distribution of the network graphs in the increasing order of their spectral radius ratio for node degree (1.01 to 3.23). The US College Football network exhibits minimal variations in the degree of its vertices (each team has more or less played against an equal number of other teams). The US Airports network exhibits maximum variation in the degree of its vertices (there are some hub airports from which there are flights to several other airports; whereas there are several airports with only fewer connections to other airports). In between these two extremes of networks, we have the other four network graphs, all of which have a spectral radius ratio for node degree around 1.4-1.7, indicating a moderate variation in the node degree (compared to the spectral radius ratios observed for the US College Football network and the US Airports network).

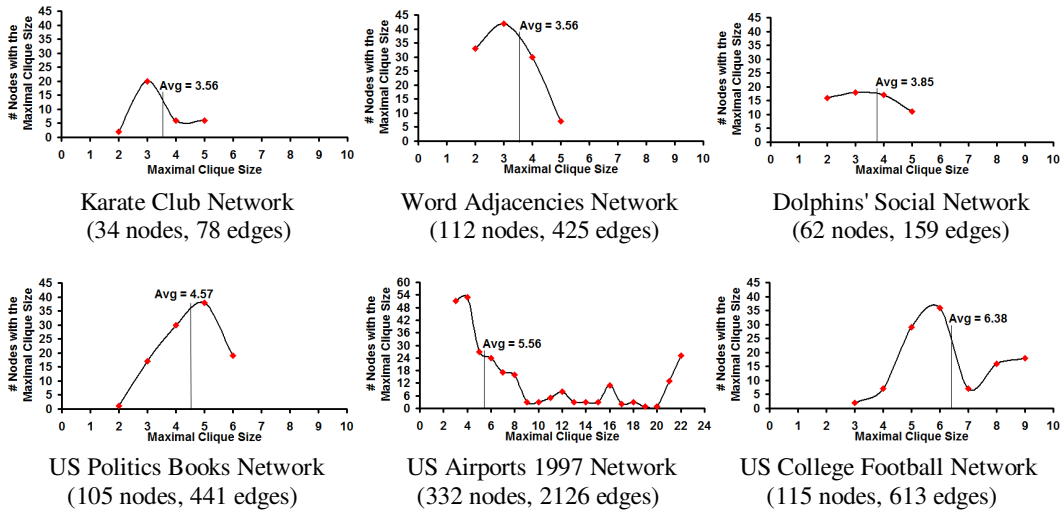

Karate Club Network
(34 nodes, 78 edges)

Word Adjacencies Network
(112 nodes, 425 edges)

Dolphins' Social Network
(62 nodes, 159 edges)

US Politics Books Network
(105 nodes, 441 edges)

US Airports 1997 Network
(332 nodes, 2126 edges)

US College Football Network
(115 nodes, 613 edges)

Figure 12. Distribution of Maximal Clique Size of the Vertices in Real-World Network Graphs

## 5.2. Maximal Clique Size Distribution of the Real-World Network Graphs

Figure 12 presents the distribution of the maximal clique size of the vertices for the six real-world network graphs, in the increasing order of the average value for the maximal clique size of the vertices. An interesting observation is that five of the six real-world network graphs exhibit a Poisson-style distribution for the maximal clique size and the average value of the maximal clique size for the nodes is very close to the maximum value. The only real-world network that does not exhibit a Poisson-style distribution for the maximal clique size is the US Airports network whose distribution of the maximal clique size appears to be more of a scale-free (power-law) pattern with a long tail (wherein the average maximal clique size is 5.56, but there exists a significant number of nodes whose maximal clique size values are 21 and 22). We can also notice that the average value of the maximal clique size of the nodes is not proportional to the number of nodes in the network nor to the spectral radius ratio for node degree. This is evident from three of the six real-world networks with comparable number of nodes (Word Adjacency Network - 112 nodes, US Politics Books Network - 112 nodes and the US Football Network - 105 nodes) incurring significantly different average values for the maximal clique size (3.56, 4.57 and 6.38 respectively). Similarly, though the spectral radius ratio for node degree increases with increase in the scale-free nature of the networks, we do not observe any such pattern of increase or decrease for the maximal clique size; for example: the US Football Network, Word Adjacency Network and the US Airports Network have spectral radius ratio for node degree values of 1.01, 1.73 and 3.22 respectively; whereas, their average maximal clique size values are 6.38, 3.56 and 5.56 respectively (no pattern of increase or decrease with increase in the spectral radius ratio for node degree).

## 6. CORRELATION COEFFICIENT ANALYSIS: MAXIMAL CLIQUE SIZE VS. CLUSTERING COEFFICIENT

The clustering coefficient of a node is the ratio of the number of links between the neighbors of the node to that of the maximum possible number of links between the neighbors of the node [1]. If a node $i$ has $k_i$ neighbors and there are $L_i$ links among these $k_i$ neighbors, then the clustering coefficient for node $i$ is: $C_i = \dfrac{L_i}{k_i(k_i-1)/2}$. In this section, we examine whether the clustering coefficient of the nodes in the six real world network graphs is positively correlated to the maximal clique size of the nodes in these graphs. Our reasoning is that a clique comprises of links between any two of its constituent nodes; thus, the neighbors of a node in a clique are also connected with links among themselves. We wanted to examine whether or not this corresponds to links between any two neighbors of a node in the real world network graph itself.

If $\overline{X}$ and $\overline{Y}$ are the average values of the two metrics (say X and Y) observed for the vertices (IDs 1 to $n$, where $n$ is the number of vertices) in the network, the formula used to compute the Correlation Coefficient between two metrics X and Y is given in equation (1), as follows:

$$CorrCoeff(X,Y) = \frac{\sum_{ID=1}^{n}(X[ID]-\overline{X})*(Y[ID]-\overline{Y})}{\sqrt{\sum_{ID=1}^{N}(X[ID]-\overline{X})^2}\sqrt{\sum_{ID=1}^{N}(Y[ID]-\overline{Y})^2}} \quad (1)$$

Table 1. Correlation Coefficient between Maximal Clique Size and Clustering Coefficient

| Network Index | Network Name | Spectral Radius Ratio for Node Degree | Correlation Coefficient: Maximal Clique Size vs. Clustering Coefficient |
|---|---|---|---|
| (vi) | US Airports 1997 Network | 3.22 | -0.47 |
| (i) | Karate Club Network | 1.46 | -0.22 |
| (ii) | Dolphins' Social Network | 1.40 | -0.17 |
| (iv) | Word Adjacencies Network | 1.73 | -0.09 |
| (iii) | US Politics Books Network | 1.41 | 0.07 |
| (v) | US College Football Network | 1.01 | 0.69 |

Table 1 lists the correlation coefficient observed for the clustering coefficient and the maximal clique size of the nodes for the six real world network graphs (in the order of increasing values of the correlation coefficient), along with the spectral radius ratio for node degree observed for these networks. Contrary to our hypothesis, we observe the clustering coefficient of the nodes in four of the six real world network graphs to be poorly correlated to the maximal clique size of the nodes; the exceptions being the US College Football network (a random network graph) and the US Airports' 97 network (a scale-free network graph) exhibiting respectively moderate levels of positive and negative correlations between the clustering coefficient and the maximal clique size of the nodes. Hence, if at all a positive correlation is observed between the clustering coefficient and maximal clique size, it is most likely by chance. On the other hand, the correlation between the clustering coefficient and maximal clique size turns more negative with increase in the scale-free nature of the networks. For networks that have moderate values of the spectral radius ratio for node degree (that is the networks are neither scale-free nor random), there is pretty much no correlation between the clustering coefficient and maximal clique size of the nodes.

## 7. CORRELATION COEFFICIENT ANALYSIS: MAXIMAL CLIQUE SIZE VS. NODE DEGREE

In this section, we present the results of correlation coefficient analysis conducted between node degree vis-a-vis the maximal size clique that each vertex is part of. The analysis has been conducted on the six real-world network graphs (mentioned in Section 5) with respect to the node degree and the maximal clique size measured for the vertices in these graphs. We implemented the exact algorithm to determine the maximal clique size for each of the vertices in a graph (see Figure 2). The visualization figures presented in the paper were obtained by porting the network data as well as the node degree/maximal clique size results to Gephi [16] and making appropriate changes to the settings in the latter. The layout algorithm chosen in Gephi for visualization is the Fruchterman Reingold algorithm [17] that presents the network in a circular format (like a globe).

Table 2. Correlation Coefficient between Maximal Clique Size and Node Degree

| Network Index | Network Name | Spectral Radius Ratio for Node Degree | Correlation Coefficient: Maximal Clique Size vs. Node Degree |
|---|---|---|---|
| (vi) | US Airports 1997 Network | 3.22 | 0.87 |
| (i) | Karate Club Network | 1.46 | 0.64 |
| (ii) | Dolphins' Social Network | 1.40 | 0.78 |
| (iv) | Word Adjacencies Network | 1.73 | 0.71 |
| (iii) | US Politics Books Network | 1.41 | 0.70 |
| (v) | US College Football Network | 1.01 | 0.32 |

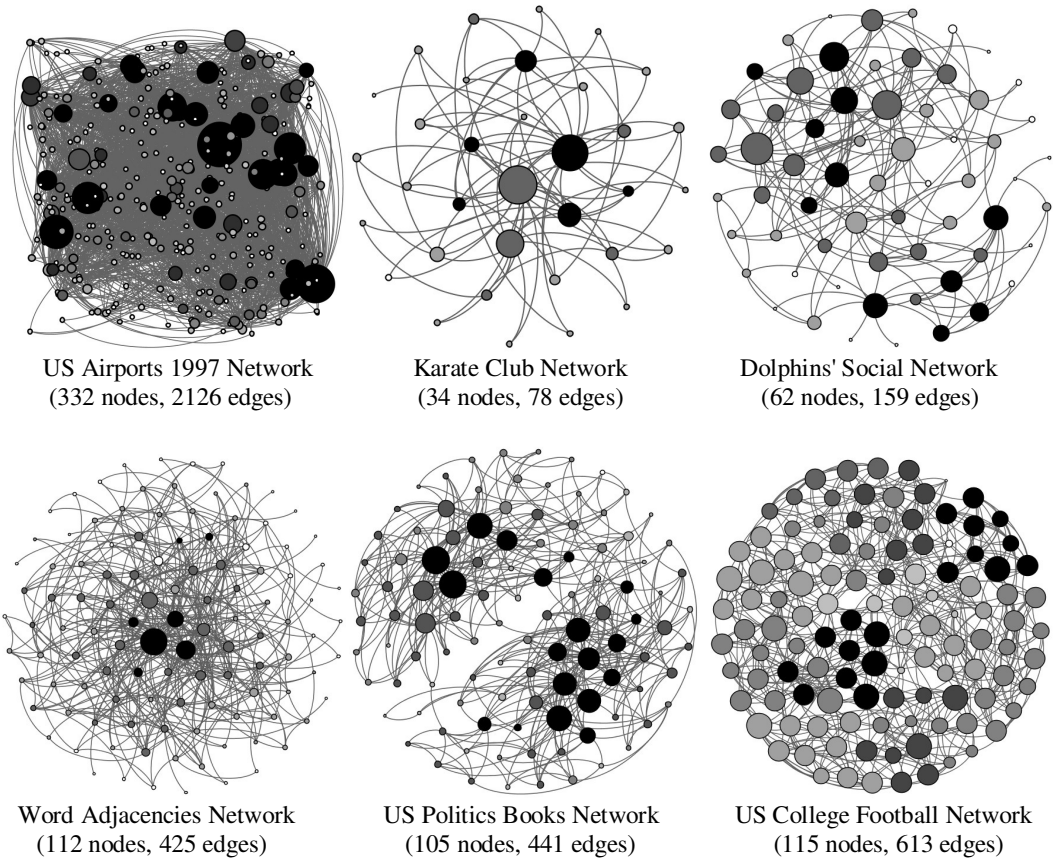

US Airports 1997 Network (332 nodes, 2126 edges)

Karate Club Network (34 nodes, 78 edges)

Dolphins' Social Network (62 nodes, 159 edges)

Word Adjacencies Network (112 nodes, 425 edges)

US Politics Books Network (105 nodes, 441 edges)

US College Football Network (115 nodes, 613 edges)

Figure 13. Correlation of Maximal Clique Size of the Vertices and Node Degree in the Real-World Network Graphs

Table 2 presents a correlation coefficient analysis of node degree and the maximal clique size observed for the vertices in each of the six real-world network graphs (in the decreasing order of the spectral radius ratio for node degree). As we can see in Table 2, in general, the correlation between the node degree and the maximal clique size increases as the spectral radius ratio for node degree increases. This implies, the more scale-free a real-world network is, the higher the correlation between the centrality value and the maximal clique size observed for a node. With several of the real-world networks being mostly scale-free, we expect these networks to exhibit a similar correlation to that observed in this paper. Also, since the correlation between the maximal clique size and node degree is the lowest (correlation coefficient of 0.31) for the US College Football Network (a random network), we conjecture that the stronger correlation (correlation coefficient of 0.7 or above) observed between these two metrics in the other five real-world network graphs is not merely by chance.

Figure 13 depicts the correlation observed for the node degree with that of the maximal clique size for the vertices in the real-world network graphs. In these figures, the node size is a measure of the node degree (the larger a node is, the larger is its degree); the node color is a measure of the maximal size clique the vertex is part of (the darker a node is, the larger is the size of the maximal clique for the node). We observe a high correlation between maximal clique size of nodes and nodes with a higher degree as well as located in a neighborhood of high degree nodes. That is, a node with high degree as well as located in a neighborhood of high degree vertices is more likely to be part of a maximal clique of larger size. In addition, as the networks get increasingly scale-free, nodes with high degree are more likely connected to other similar nodes with high degree (to facilitate an average path length that is almost independent of network size: characteristic of scale-free networks [1]) contributing to a positive correlation between degree-based centrality metrics and maximal clique size. We anticipate that as the networks become increasingly scale-free, the hubs (that facilitate shortest-path communication between any two nodes) are more likely to form the maximum clique for the entire network graph - contributing to higher levels of positive correlation between node degree and maximal clique size.

## 8. ASSORTATIVITY INDEX-BASED ANALYSIS: MAXIMAL CLIQUE SIZE AND NODE DEGREE

The assortativity index for a network graph with respect to a particular node-related metric is a measure of the association of nodes with similar values for the metric [1]. For example, the assortativity index of a graph with respect to node degree is a measure of the association of higher degree nodes with other high degree nodes as well as the association of nodes of lower degree nodes with other lower degree nodes. In this section, we analyze the assortativity index of the six real-world network graphs with respect to the maximal clique size and node degree, and examine the nature of association between nodes having higher values for each of these two metrics. If $m$ is the node-related metric of interest, then the assortativity index of the network graph with respect to $m$ is evaluated as the correlation coefficient of the values (with respect to metric $m$) for the end nodes of the edges in the graph. Consider a network graph of $n$ nodes and set of undirected (bi-directional) edges $E$; let $m[1], m[2], ...., m[n]$ be the values for nodes 1, 2, ...,$n$ with respect to metric $m$ and $\overline{m}$ be the average value of the metric, the assortativity index with respect to metric $m$ is given by equation (2).

$$AssortativityIndex(m) = \frac{\sum_{(i,j) \in E} (m[i] - \overline{m}) * (m[j] - \overline{m})}{\sqrt{\sum_{(i,j) \in E} (m[i] - \overline{m})^2} \sqrt{\sum_{(i,j) \in E} (m[j] - \overline{m})^2}} \quad (2)$$

Positive values for the assortativity index with respect to a metric indicates that the network exhibits assortativity with respect to the metric (nodes with higher values for the metric are more likely to be connected to nodes with higher values for the metric and vice-versa); negative values for the assortativity index indicates the network exhibits disassortativity (nodes with lower values for the metric are more likely to be connected to nodes with higher values for the metric and vice-versa); assortativity index values close to 0 indicates the network is more neutral with respect to the metric (i.e., the values for the end nodes of the edges with respect to the metric do not exhibit any correlation).

Table 3 lists the assortativity index values for the maximal clique size and degree of the vertices for the six real-world network graphs, along with their spectral radius ratio for node degree. We observe the assortativity index (with respect to the maximal clique size) for all the six network graphs to be positive and the assortativity index value for the maximal clique size increases with increase in the level of randomness in the network, indicating that the association of nodes of a particular maximal clique size with other nodes that are also of the same maximal clique size is more by chance. On the other hand, we observe the assortativity index (with respect to the node degree) for five of the six network graphs (i.e., all network graphs, except the US Football Network that exhibits the characteristic of random graphs) to be negative and the assortativity index values for the node degree gets more negative with increase in the scale-free nature of the network, indicating high degree nodes are more likely to be associated with low degree nodes (especially with increase in the spectral radius ratio for node degree). Finally, to confirm our earlier observation of a positive correlation between maximal clique size of the vertices and node degree, we observe in Table 3 that the six-real world networks could be listed in an identical order, in the increasing order of the Assortativity Index of the network graphs with respect to both maximal clique size of the vertices and node degree.

Table 3. Assortativity Index of the Real-World Network Graphs based on Maximal Clique Size of the Vertices and Node Degree

| *Network Index* | *Network Name* | *Spectral Radius Ratio for Node Degree* | *Assortativity Index for Maximal Clique Size* | *Assortativity Index for Node Degree* |
|---|---|---|---|---|
| (i) | Zachary's Karate Network | 1.46 | 0.13 | -0.48 |
| (vi) | US Airports 1997 Network | 3.22 | 0.17 | -0.21 |
| (iv) | Word Adjacencies Network | 1.73 | 0.20 | -0.09 |
| (iii) | US Politics Books Network | 1.41 | 0.20 | -0.04 |
| (ii) | Dolphins' Social Network | 1.40 | 0.23 | -0.02 |
| (v) | US College Football Network | 1.01 | 0.59 | 0.19 |

## 9. CONCLUSIONS

The first half of the paper reveals interesting observations with regards to the distribution of the maximal clique size per node for small-world networks and random networks that evolve from a regular network. As we transform from a regular network (with $K_{regular}$ number of links per node) to a small-world network through link rewiring, we observe the maximal clique size of the nodes to be invariant and very close to that of the average maximal clique size per node as well as close to that of the average maximal clique size per node in the regular network. As we transform from a small-world network to a random network (by increasing the probability of rewiring), we observe the distribution of the maximal clique size per node to become more broader and thereby the probability of observing a maximal clique size per node close to that of the average maximal clique size is relatively much lower. Also, with increase in the probability of rewiring, the distributions for the maximal clique size obtained for different $K_{regular}$ values overlap each other and shift towards a lower average value. Nevertheless, for all the scenarios/values for the probability of rewiring and the initial number of links per node, the distribution for the maximal clique size reflects that of a Poisson distribution.

Similar to that of the theoretically simulated networks, in the second half of the paper, we also observe a Poisson-style distribution for maximal clique size of the vertices in real-world network graphs irrespective of the number of nodes and the degree distribution of the vertices is an interesting observation that has not been hitherto reported in the literature. We conjecture the distribution for the maximal clique size of the vertices to transform from Poisson to Power-law distribution in networks that are highly scale-free (as observed in the case of the US Airports'97 Network). With the problem of determining maximal clique sizes for individual vertices being computationally time consuming, our approach taken in this paper to study the correlation between maximal clique sizes vis-a-vis node degree and clustering coefficient could be the first step in identifying correlation between cliques/clique size in real-world network graphs and one or more computationally-light node-based network metrics that can be quickly determined and henceforth appropriate inferences can be made about a ranking of the vertices with respect to maximal clique size. The approach taken to first to find the correlation coefficient between the two metrics of interest (like node degree and maximal clique size of the vertices) in the individual network graphs and then ranking the network graphs in the increasing order of the Assortativity Index of the graphs with respect to each of the two metrics (an identical or close to identical listing of the network graphs with respect to the each of the two metrics vindicates the positive correlation observed between the two metrics based on correlation coefficient analysis). Such an approach for correlation study between two node-based metrics is a unique approach that has been so far not presented in the literature. We observe node degree to show promising positive correlations to that of maximal clique sizes of the individual vertices, especially as the networks get increasingly scale-free; this observation could form the basis of future research for analysis of maximal clique size of the vertices in complex real-world network graphs and the correlations of the maximal clique size of the vertices with other computationally-light metrics for complex network analysis.

**Author**

Dr. Natarajan Meghanathan is a Full Professor of Computer Science at Jackson State University, USA. His areas of research interests are Network Science and Graph Theory, Wireless Ad hoc Networks and Sensor Networks, Cyber Security and Machine Learning. He has published more than 150 peer-reviewed articles and obtained grants from several federal agencies. He serves as the editor-in-chief of three international journals as well as serves in the organizing committees of several international conferences.

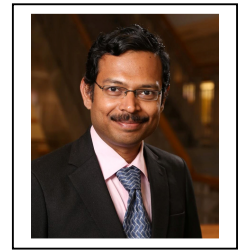